\documentclass{article}

\usepackage{arxiv}

\usepackage[utf8]{inputenc} % allow utf-8 input
\usepackage[T1]{fontenc}    % use 8-bit T1 fonts
\usepackage{lmodern}        % https://github.com/rstudio/rticles/issues/343
\usepackage{hyperref}       % hyperlinks
\usepackage{url}            % simple URL typesetting
\usepackage{booktabs}       % professional-quality tables
\usepackage{amsfonts}       % blackboard math symbols
\usepackage{nicefrac}       % compact symbols for 1/2, etc.
\usepackage{microtype}      % microtypography
\usepackage{graphicx}

\title{Introducing ICBe: Very High Recall and Precision Event Extraction
from Narratives about International Crises}

\author{
    Rex W. Douglass
   \\
    University of California, San Diego \\
   \\
  \texttt{} \\
   \And
    Thomas Leo Scherer
   \\
    University of California, San Diego \\
   \\
  \texttt{} \\
   \And
    J. Andrés Gannon
   \\
    Vanderbilt University \\
   \\
  \texttt{} \\
   \And
    Erik Gartzke
   \\
    University of California, San Diego \\
   \\
  \texttt{} \\
   \And
    Jon Lindsay
   \\
    Georgia Institute of Technology \\
   \\
  \texttt{} \\
   \And
    Shannon Carcelli
   \\
    University of Maryland \\
   \\
  \texttt{} \\
   \And
    Jonathan Wilkenfeld
   \\
    University of Maryland \\
   \\
  \texttt{} \\
   \And
    David M. Quinn
   \\
    University of Maryland \\
   \\
  \texttt{} \\
   \And
    Catherine Aiken
   \\
    Georgetown University \\
   \\
  \texttt{} \\
   \And
    Jose Miguel Cabezas Navarro
   \\
    Universidad Mayor \\
   \\
  \texttt{} \\
   \And
    Neil Lund
   \\
    University of Maryland \\
   \\
  \texttt{} \\
   \And
    Egle Murauskaite
   \\
    University of Maryland \\
   \\
  \texttt{} \\
   \And
    Diana Partridge
   \\
    University of Maryland \\
   \\
  \texttt{} \\
  }

\newlength{\cslhangindent}
\newlength{\csllabelwidth}
\newlength{\cslentryspacingunit} % times entry-spacing
  {}%
  {\par}
\newenvironment{CSLReferences}[2] % #1 hanging-ident, #2 entry spacing
 {% don't indent paragraphs
  \setlength{\parindent}{0pt}
  \ifodd #1
  \let\oldpar\par
  \def\par{\hangindent=\cslhangindent\oldpar}
  \fi
  \setlength{\parskip}{#2\cslentryspacingunit}
 }%
 {}
\usepackage{calc}

\newcommand{\CSLLeftMargin}[1]{\parbox[t]{\csllabelwidth}{#1}}
\newcommand{\CSLRightInline}[1]{\parbox[t]{\linewidth - \csllabelwidth}{#1}\break}

\usepackage[utf8]{inputenc}
\usepackage{pifont}
\usepackage{newunicodechar}
\newunicodechar{✓}{\ding{51}}
\newunicodechar{✗}{\ding{55}}
\usepackage{array}
\usepackage{ctable}
\usepackage{booktabs}
\usepackage{longtable}
\usepackage{array}
\usepackage{multirow}
\usepackage{wrapfig}
\usepackage{colortbl}
\usepackage{pdflscape}
\usepackage{tabu}
\usepackage{threeparttable}
\usepackage{threeparttablex}
\usepackage[normalem]{ulem}
\usepackage{makecell}
\usepackage{titlesec}
\usepackage[parfill]{parskip}
\usepackage{makecell}
\usepackage{graphicx}
\usepackage{setspace}
\usepackage{cellspace}
\renewcommand{\arraystretch}{0.8}
\AtBeginEnvironment{lltable}{\singlespacing}
\usepackage{subcaption}
\usepackage{atbegshi}% http://ctan.org/pkg/atbegshi
\usepackage{float}
\usepackage[algo2e]{algorithm2e}
\usepackage{multirow}
\usepackage{multicol}
\usepackage{colortbl}
\usepackage{hhline}
\usepackage{longtable}
\usepackage{array}
\usepackage{hyperref}
\begin{document}
\maketitle

\begin{abstract}
How do international crises unfold? We conceptualize of international
relations as a strategic chess game between adversaries and develop a
systematic way to measure pieces, moves, and gambits accurately and
consistently over a hundred years of history. We introduce a new
ontology and dataset of international events called ICBe based on a very
high-quality corpus of narratives from the International Crisis Behavior
(ICB) Project. We demonstrate that ICBe has higher coverage, recall, and
precision than existing state of the art datasets and conduct two
detailed case studies of the Cuban Missile Crisis (1962) and
Crimea-Donbas Crisis (2014). We further introduce two new event
visualizations (event icongraphy and crisis maps), an automated
benchmark for measuring event recall using natural language processing
(sythnetic narratives), and an ontology reconstruction task for
objectively measuring event precision. We make the data, online
appendix, replication material, and visualizations of every historical
episode available at a companion website www.crisisevents.org and the
github repository.
\end{abstract}


\twocolumn

If we could record every important interaction between countries in all
of diplomacy, military conflict, and international political economy,
how much unique information would this chronicle amount to, and how
surprised would we be to see something new? In other words, what is the
entropy of international relations? This record could in principle be
unbounded, but the central conceit of social science is that there are
structural regularities that limit what actors can do, their best
options, and even which actors are likely to survive (1, 2). If so, then
these events can be systematically measured, and accordingly, massive
effort is expended in social science attempting to record these
regularities.\footnote{See work on crises (3, 4), militarized disputes
  (5--7), wars (8, 9), organized violence (10, 11), political violence
  (12), sanctions (13), trade (14), and international agreements
  (15--17), dispute resolution (17, 18), and diplomacy (19, 20).} Thanks
to improvements in natural language processing, more open-ended efforts
have begun to capture entire unstructured streams of events from
international news reports.\footnote{See (21); (22); (23); (24); (25);
  (26). On event-extraction from images and social-media see (27) and
  (28).} How close these efforts are to accurately measuring all or even
most of what is essential in international relations is an open
empirical question, one for which we provide new evidence here.

Our contribution is a high coverage ontology and event dataset for key
historical episodes in 20th and 21st-century international relations. We
develop a large, flexible ontology of international events with the help
of both human coders and natural language processing. We apply it
sentence-by-sentence to an unusually high-quality corpus of historical
narratives of international crises (1, 29--32). The result is a new
lower bound estimate of how much actually happens between states during
pivotal historical episodes. We then develop several methods for
objectively gauging how well these event codings reconstruct the
information contained in the original narrative. We conclude by
benchmarking our event codings against several current state-of-the-art
event data collection efforts. We find that existing systems produce
sequences of events that do not contain enough information to
reconstruct the underlying historical episode. The underlying
fine-grained variation in international affairs is unrecognizable
through the lens of current quantification efforts.

This is a measurement paper that makes the following argument --- there
is a real-world unobserved latent concept known as international
relations, we propose a method for systematically measuring it, we
successfully apply this method producing a new large scale set of
measurements, those measurements exhibit several desirable kinds of
internal and external validity, and those measurements out-perform other
existing approaches. The article organizes that argument into eight
sections: task definition; corpus; priors/existing state of the art;
ICBe coding process; internal consistency; case study selection; recall;
and precision. A final section concludes.

\hypertarget{task-definition}{%
\section*{Task Definition}\label{task-definition}}
\addcontentsline{toc}{section}{Task Definition}

We consider the measurement task of abstracting discrete events about a
historical episode in international relations. The easiest way to convey
the task is with an example. Figure 1 shows a narrative account of the
Cuban Missile Crisis (1962) alongside a mapping from each natural
language sentence to discrete machine readable abstractive events.
Formally, a historical episode, \(H\), is demarcated by a period of time
\([T_{start}, T_{end}] \in T\), a set of Players \(p \in P\), and a set
of behaviors they undertook during that time \(b \in B\). International
Relations, \(IR\), is the system of regularities that govern the
strategic interactions that world actors make during a historical
episode, given their available options, preferences, beliefs, and
expectations of choices made by others. We observe neither \(H\) nor
\(IR\) directly. Rather the Historical Record, \(HR\), produces
documents \(d \in D\) containing some relevant and true (as well as
irrelevant and untrue) information about behaviors that were undertaken
recorded in the form of unstructured natural language text. The task is
to combine informative priors about \(IR\) with an unstructured corpus
\(D\) to produce a series of structured discrete events, \(e \in E\),
that have high coverage, precision, and recall over what actually took
place in history, \(H\).

\clearpage
\onecolumn
\begin{figure}[H]
\caption{Case Study 1: Cuban Missile Crisis (1962) - ICB Narrative vs. ICBe Events \label{fig:case_study_cuban_precision}}
\centering{\includegraphics[ height=19cm]{"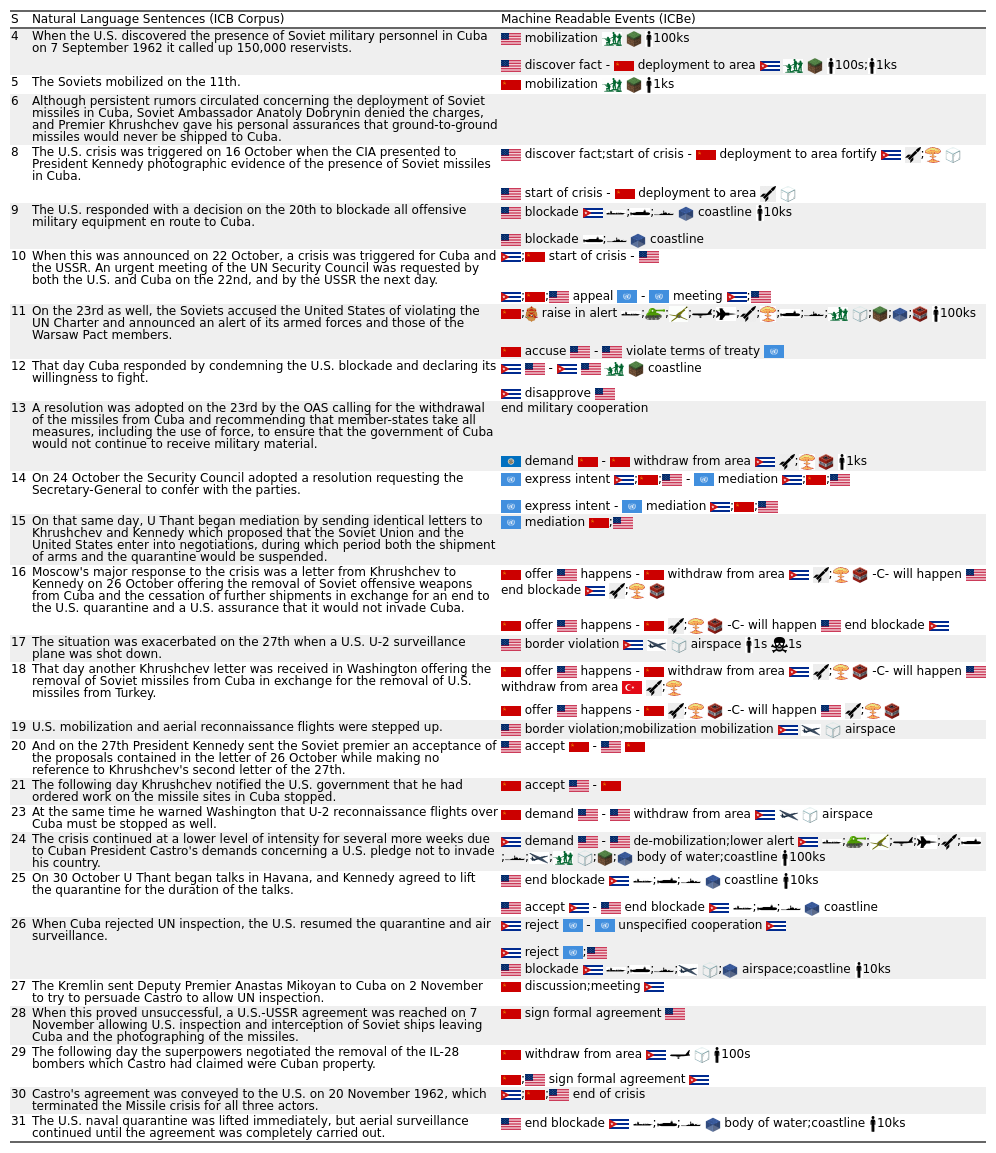"}}
\end{figure}
\clearpage
\twocolumn

\hypertarget{corpus}{%
\section*{Corpus}\label{corpus}}
\addcontentsline{toc}{section}{Corpus}

For our corpus, \(D\), we select a set of unusually high-quality
historical narratives from the International Crisis Behavior (ICB)
project (\(n=471\))(SI Appendix, Table A1)(33, 34).\footnote{The Online
  Appendix is at the
  \href{https://urldefense.com/v3/__https://github.com/CenterForPeaceAndSecurityStudies/ICBEventData__;!!Mih3wA!WxDJtEczKfxGTh0S2Krunap8ReymFEL5iTWaSfOHeqlSdyfRx77zmjBSWO1OAm13$}{ICBEventData
  Github Repository}.} Their domain is 20th and 21st-century crises,
defined as a change in the type, or an increase in the intensity, of
disruptive interaction with a heightened probability of military
hostilities that destabilizes states' relationships or challenges the
structure of the international system (3).\footnote{On near crises see
  (32).} Crises are a significant focus of detailed single case studies
and case comparisons because they provide an opportunity to examine
behaviors in IR short of, or at least prior to, full conflict (3,
35--42). Case selection was exhaustive based on a survey of world news
archives and region experts, cross-checked against other databases of
war and conflict, and non-English sources (33, 43). Each narrative was
written by consensus by a small number of scholars, using a uniform
coding scheme, with similar specificity (44). The corpus is unique in IR
because it is designed to be used in a downstream quantitative coding
project.

\hypertarget{prior-beliefs-about-ir-ontological-coverage-and-the-existing-state-of-the-art}{%
\section*{Prior Beliefs about IR, Ontological Coverage, and the Existing
State of the
Art}\label{prior-beliefs-about-ir-ontological-coverage-and-the-existing-state-of-the-art}}
\addcontentsline{toc}{section}{Prior Beliefs about IR, Ontological
Coverage, and the Existing State of the Art}

Next we draw informative prior beliefs about the underlying process of
IR that we expect to govern behavior during historical episodes and
their conversion to the historical record. We organize our prior beliefs
along two overarching axes, summarized in detail by Table 1.

The first axis (rows) represents the types of information we expect to
find in IR and forms the basis for our proposed ontology. We employ a
metaphor of a chess game, with players (polities, rebel groups, IGOs,
etc.), pieces (military platforms, civilians, domains), and behaviors
(think, say, do). Precise sequencing is required to capture gambits
(sequences of moves) and outcomes (victory, defeat, peace, etc.), while
precise geo-coding is required to understand the chessboard (medium of
conflict). We find 472 actors and 117 different behaviors, and provide a
full codebook in the online material.\footnote{See the Github Repository
  \href{https://urldefense.com/v3/__https://github.com/CenterForPeaceAndSecurityStudies/ICBEventData__;!!Mih3wA!WxDJtEczKfxGTh0S2Krunap8ReymFEL5iTWaSfOHeqlSdyfRx77zmjBSWO1OAm13$}{ICBEventData}.}

We base our informed priors primarily on two sources of information. The
first is the extensive existing measurement efforts of IR which we
provide citations to alongside each concept. Second, we performed
preliminary natural language processing of the corpus and identified
named entities and behaviors mentioned in the text. Verbs were matched
to the most likely definition found in Wordnet (45), tallied, and then
aggregated into a smaller number hypernyms balancing conceptual detail
and manageable sparsity for human coding (SI Appendix, Table A2).

The second axis (columns) compares the very high ontological coverage of
ICBe to existing state of the art systems in production and with global
coverage. They begin with our contribution ICBe, alongside other
event-level datasets including CAMEO dictionary lookup based systems
(Historical Phoenix (46); ICEWS (24, 25); Terrier (26)), the Militarized
Interstate Disputes Incidents dataset, and the UCDP-GED dataset (10, 11,
47).\footnote{Additional relevant but dated or too small of an overlap
  in domain include BCOW (48), WEIS (49), CREON (50), CASCON (51),
  SHERFACS (52), Real-Time Phoenix (23), and COfEE (53) (see histories
  in (54) and (55)).} The final set of columns compares episode-level
datasets beginning with the original ICB project (3, 4, 56); the
Militarized Interstate Disputes dataset (5, 6, 57, 58), and the
Correlates of War (8). With the exception of large scale CAMEO
dictionary based systems, the existing state of the art quantitative
datasets ignore the vast majority of the information content found in
international relations.\footnote{See (53) for a recent review of
  ontological depth and availability of Gold Standard example text.}

\clearpage
\onecolumn

{[}1{]} 37 17

\providecommand{\docline}[3]{\noalign{\global\setlength{\arrayrulewidth}{#1}}\arrayrulecolor[HTML]{#2}\cline{#3}}

\setlength{\tabcolsep}{2pt}

\renewcommand*{\arraystretch}{0.75}

\begin{longtable}[c]{|p{0.10in}|p{0.10in}|p{3.00in}|p{0.25in}|p{0.25in}|p{0.25in}|p{0.25in}|p{0.25in}|p{0.25in}|p{0.25in}|p{0.25in}|p{0.25in}}

\caption{Ontological coverage of ICBe versus existing State of the Art
}\\

\hhline{>{\arrayrulecolor[HTML]{000000}\global\arrayrulewidth=2pt}->{\arrayrulecolor[HTML]{000000}\global\arrayrulewidth=2pt}->{\arrayrulecolor[HTML]{000000}\global\arrayrulewidth=2pt}->{\arrayrulecolor[HTML]{000000}\global\arrayrulewidth=2pt}->{\arrayrulecolor[HTML]{000000}\global\arrayrulewidth=2pt}->{\arrayrulecolor[HTML]{000000}\global\arrayrulewidth=2pt}->{\arrayrulecolor[HTML]{000000}\global\arrayrulewidth=2pt}->{\arrayrulecolor[HTML]{000000}\global\arrayrulewidth=2pt}->{\arrayrulecolor[HTML]{000000}\global\arrayrulewidth=2pt}->{\arrayrulecolor[HTML]{000000}\global\arrayrulewidth=2pt}->{\arrayrulecolor[HTML]{000000}\global\arrayrulewidth=2pt}->{\arrayrulecolor[HTML]{000000}\global\arrayrulewidth=2pt}-}

\multicolumn{3}{!{\color[HTML]{000000}\vrule width 2pt}>{\centering}p{\dimexpr 3.2in+4\tabcolsep+2\arrayrulewidth}}{\rotatebox[origin=c]{270}{\fontsize{7}{7}\selectfont{\textcolor[HTML]{000000}{}}}} & \multicolumn{6}{!{\color[HTML]{666666}\vrule width 1pt}>{\centering}p{\dimexpr 1.5in+10\tabcolsep+5\arrayrulewidth}}{\fontsize{7}{7}\selectfont{\textcolor[HTML]{000000}{Events\ Datasets}}} & \multicolumn{3}{!{\color[HTML]{666666}\vrule width 1pt}>{\centering}p{\dimexpr 0.75in+4\tabcolsep+2\arrayrulewidth}!{\color[HTML]{000000}\vrule width 2pt}}{\fontsize{7}{7}\selectfont{\textcolor[HTML]{000000}{Episodes\ Datasets}}} \\

\multicolumn{2}{!{\color[HTML]{000000}\vrule width 2pt}>{\centering}p{\dimexpr 0.2in+2\tabcolsep+1\arrayrulewidth}}{\rotatebox[origin=c]{270}{\fontsize{7}{7}\selectfont{\textcolor[HTML]{000000}{}}}} & \multicolumn{1}{!{\color[HTML]{000000}\vrule width 0pt}>{\centering}p{\dimexpr 3in+0\tabcolsep+0\arrayrulewidth}}{\rotatebox[origin=c]{270}{\fontsize{7}{7}\selectfont{\textcolor[HTML]{000000}{Concept}}}} & \multicolumn{1}{!{\color[HTML]{666666}\vrule width 1pt}>{\centering}p{\dimexpr 0.25in+0\tabcolsep+0\arrayrulewidth}}{\rotatebox[origin=c]{270}{\fontsize{7}{7}\selectfont{\textcolor[HTML]{000000}{ICBe}}}} & \multicolumn{1}{!{\color[HTML]{000000}\vrule width 0pt}>{\centering}p{\dimexpr 0.25in+0\tabcolsep+0\arrayrulewidth}}{\rotatebox[origin=c]{270}{\fontsize{7}{7}\selectfont{\textcolor[HTML]{000000}{Phoenix}}}} & \multicolumn{1}{!{\color[HTML]{000000}\vrule width 0pt}>{\centering}p{\dimexpr 0.25in+0\tabcolsep+0\arrayrulewidth}}{\rotatebox[origin=c]{270}{\fontsize{7}{7}\selectfont{\textcolor[HTML]{000000}{Terrier}}}} & \multicolumn{1}{!{\color[HTML]{000000}\vrule width 0pt}>{\centering}p{\dimexpr 0.25in+0\tabcolsep+0\arrayrulewidth}}{\rotatebox[origin=c]{270}{\fontsize{7}{7}\selectfont{\textcolor[HTML]{000000}{ICEWs}}}} & \multicolumn{1}{!{\color[HTML]{000000}\vrule width 0pt}>{\centering}p{\dimexpr 0.25in+0\tabcolsep+0\arrayrulewidth}}{\rotatebox[origin=c]{270}{\fontsize{7}{7}\selectfont{\textcolor[HTML]{000000}{MIDs\ Incidents}}}} & \multicolumn{1}{!{\color[HTML]{000000}\vrule width 0pt}>{\centering}p{\dimexpr 0.25in+0\tabcolsep+0\arrayrulewidth}}{\rotatebox[origin=c]{270}{\fontsize{7}{7}\selectfont{\textcolor[HTML]{000000}{UCDP-GED}}}} & \multicolumn{1}{!{\color[HTML]{666666}\vrule width 1pt}>{\centering}p{\dimexpr 0.25in+0\tabcolsep+0\arrayrulewidth}}{\rotatebox[origin=c]{270}{\fontsize{7}{7}\selectfont{\textcolor[HTML]{000000}{ICB}}}} & \multicolumn{1}{!{\color[HTML]{000000}\vrule width 0pt}>{\centering}p{\dimexpr 0.25in+0\tabcolsep+0\arrayrulewidth}}{\rotatebox[origin=c]{270}{\fontsize{7}{7}\selectfont{\textcolor[HTML]{000000}{MIDs}}}} & \multicolumn{1}{!{\color[HTML]{000000}\vrule width 0pt}>{\centering}p{\dimexpr 0.25in+0\tabcolsep+0\arrayrulewidth}!{\color[HTML]{000000}\vrule width 2pt}}{\rotatebox[origin=c]{270}{\fontsize{7}{7}\selectfont{\textcolor[HTML]{000000}{COW}}}} \\

\hhline{>{\arrayrulecolor[HTML]{000000}\global\arrayrulewidth=2pt}->{\arrayrulecolor[HTML]{000000}\global\arrayrulewidth=2pt}->{\arrayrulecolor[HTML]{000000}\global\arrayrulewidth=2pt}->{\arrayrulecolor[HTML]{000000}\global\arrayrulewidth=2pt}->{\arrayrulecolor[HTML]{000000}\global\arrayrulewidth=2pt}->{\arrayrulecolor[HTML]{000000}\global\arrayrulewidth=2pt}->{\arrayrulecolor[HTML]{000000}\global\arrayrulewidth=2pt}->{\arrayrulecolor[HTML]{000000}\global\arrayrulewidth=2pt}->{\arrayrulecolor[HTML]{000000}\global\arrayrulewidth=2pt}->{\arrayrulecolor[HTML]{000000}\global\arrayrulewidth=2pt}->{\arrayrulecolor[HTML]{000000}\global\arrayrulewidth=2pt}->{\arrayrulecolor[HTML]{000000}\global\arrayrulewidth=2pt}-}

\endfirsthead

\hhline{>{\arrayrulecolor[HTML]{000000}\global\arrayrulewidth=2pt}->{\arrayrulecolor[HTML]{000000}\global\arrayrulewidth=2pt}->{\arrayrulecolor[HTML]{000000}\global\arrayrulewidth=2pt}->{\arrayrulecolor[HTML]{000000}\global\arrayrulewidth=2pt}->{\arrayrulecolor[HTML]{000000}\global\arrayrulewidth=2pt}->{\arrayrulecolor[HTML]{000000}\global\arrayrulewidth=2pt}->{\arrayrulecolor[HTML]{000000}\global\arrayrulewidth=2pt}->{\arrayrulecolor[HTML]{000000}\global\arrayrulewidth=2pt}->{\arrayrulecolor[HTML]{000000}\global\arrayrulewidth=2pt}->{\arrayrulecolor[HTML]{000000}\global\arrayrulewidth=2pt}->{\arrayrulecolor[HTML]{000000}\global\arrayrulewidth=2pt}->{\arrayrulecolor[HTML]{000000}\global\arrayrulewidth=2pt}-}

\multicolumn{3}{!{\color[HTML]{000000}\vrule width 2pt}>{\centering}p{\dimexpr 3.2in+4\tabcolsep+2\arrayrulewidth}}{\rotatebox[origin=c]{270}{\fontsize{7}{7}\selectfont{\textcolor[HTML]{000000}{}}}} & \multicolumn{6}{!{\color[HTML]{666666}\vrule width 1pt}>{\centering}p{\dimexpr 1.5in+10\tabcolsep+5\arrayrulewidth}}{\fontsize{7}{7}\selectfont{\textcolor[HTML]{000000}{Events\ Datasets}}} & \multicolumn{3}{!{\color[HTML]{666666}\vrule width 1pt}>{\centering}p{\dimexpr 0.75in+4\tabcolsep+2\arrayrulewidth}!{\color[HTML]{000000}\vrule width 2pt}}{\fontsize{7}{7}\selectfont{\textcolor[HTML]{000000}{Episodes\ Datasets}}} \\
\end{longtable}

\clearpage
\twocolumn

\hypertarget{icbe-coding-process}{%
\section*{ICBe Coding Process}\label{icbe-coding-process}}
\addcontentsline{toc}{section}{ICBe Coding Process}

The ICBe ontology follows a hierarchical design philosophy where a
smaller number of significant decisions are made early on and then
progressively refined into more specific details (59).\footnote{This
  process quickly focuses the coder on a smaller number of relevant
  options while also allowing them to apply multiple tags if the
  sentence explicitly includes more than one or there is insufficient
  evidence to choose only one tag. The guided coding process also allows
  for the possibility that earlier coarse decisions have less error than
  later fine-grained decisions.} Each coder was instructed to first
thoroughly read the full crisis narrative and then presented with a
custom graphical user interface (SI Appendix, Fig. B1). Coders then
proceeded sentence by sentence, choosing the number of events (0-3) that
occurred, the highest behavior (thought, speech, or activity), a set of
players (\(P\)), whether the means were primarily armed or unarmed,
whether there was an increase or decrease in aggression
(uncooperative/escalating or cooperative/de-escalating), and finally one
or more non-mutually exclusive specific activities. Some additional
details like location and timing information was always collected while
other details were only collected if appropriate, e.g.~force size,
fatalities, domains, units, etc. A unique feature of the ontology is
that thought, speech, and do behaviors can be nested into combinations,
e.g.~an offer for the U.S.S.R. to remove missiles from Cuba in exchange
for the U.S. removing missiles from Turkey. Through compounding, the
ontology can capture what players were said to have known, learned, or
said about other specific fully described actions.

Each crisis was typically assigned to 2 expert coders and 2 novice
coders with an additional tie-breaking expert coder assigned to
sentences with high disagreement.\footnote{Expert coders were graduate
  students or postgraduates who collaboratively developed the ontology
  and documentation for the codebook. Undergraduate coders were students
  who engaged in classroom workshops.} For the purposes of measuring
intercoder agreement and consensus, we temporarily disaggregate the unit
of analysis to the Coder-Crisis-Sentence-Tag (n=993,740), where a tag is
any unique piece of information a coder can associate with a sentence
such as an actor, date, behavior, etc. We then aggregate those tags into
final events (n=18,783), using a consensus procedure (SI Appendix,
Algorithm B2) that requires a tag to have been chosen by at least one
expert coder and either a majority of expert or novice coders. This
screens noisy tags that no expert considered possible but leverages
novice knowledge to tie-break between equally plausible tags chosen by
experts.

\hypertarget{internal-consistency}{%
\section*{Internal Consistency}\label{internal-consistency}}
\addcontentsline{toc}{section}{Internal Consistency}

We evaluate the internal validity of the coding process in several ways.
For every tag applied we calculate the observed intercoder agreement as
the percent of other coders who also applied that same tag (SI Appendix,
Fig. B3). Across all concepts, the Top 1 Tag Agreement was low among
novices (31\%), moderate for experts (65\%), and high (73\%) following
the consensus screening procedure.

We attribute the remaining disagreement primarily to three sources.
First, we required coders to rate their confidence which was observed to
be low for 20\% of sentences- half due to a mismatch between the
ontology and the text (``survey doesn't fit event''-45\%) and half due
to a lack of information or confused writing in the source text (``more
knowledge needed''-40\%, ``confusing sentence''-6\%). Observed
disagreement varied predictably with self reported confidence (SI
Appendix, Fig. B4). Second, as intended agreement is higher (75-80\%)
for questions with fewer options near the root of the ontology compared
to agreement for questions near the leafs of the ontology (50\%-60\%).
Third, individual coders exhibiting nontrivial coding styles, e.g.~some
more expressive applying many tags per concept while others focused on
only the single best match. We further observed unintended synonymity,
e.g.~the same information can be framed as either a threat to do
something or a promise not to do something.

\hypertarget{case-study-selection}{%
\section*{Case Study Selection}\label{case-study-selection}}
\addcontentsline{toc}{section}{Case Study Selection}

The remaining two qualities we seek to measure are recall and precision
of ICBe events in absolute terms and relative to other existing systems.
We provide full ICB narratives, ICBe coding in an easy to read
icongraphic form, and a wide range of visualizations for every case on
the companion website. In this paper, we focus on two deep case studies.
The first is the Cuban Missile Crisis (Figure 1) which took place
primarily in the second half of 1962, involved the United States, the
Soviet Union, and Cuba, and is widely known for bringing the world to
the brink of nuclear war (hereafter Cuban Missiles). The second is the
Crimea-Donbas Crisis (SI Appendix Figure D1) which took place primarily
in 2014, involved Russia, Ukraine, and NATO, and within a decade
spiraled into a full scale invasion (herafter Crimea-Donbas). Both cases
involve a superpower in crisis with a neighbor, initiated by a change
from a friendly to hostile regime, with implications for economic and
military security for the superpower, risked full scale invasion, and
eventually invited intervention by opposing superpowers. We choose these
cases because they are substantively significant to 20th and 21st
century international relations, widely known across scientific
disciplines and popular culture, and are sufficiently brief to evaluate
in depth.

\hypertarget{recall}{%
\section*{Recall}\label{recall}}
\addcontentsline{toc}{section}{Recall}

Recall measures the share of desired information recovered by a sequence
of coded events, \(Pr(E|H)\), and is poorly defined for historical
episodes. First, there is no genuine ground truth about what occurred,
only surviving texts about it. Second, there is no \textit{a priori}
guide to what information is necessary detail and what is ignorable
trivia. History suffers from what is known as the Coastline Paradox (60)
--- it has a fractal dimension greater than one such that the more you
zoom in the more detail you will find about individual events and in
between every two discrete events. The ICBe ontology is a proposal about
what information is important, but we need an independent benchmark to
evaluate whether that proposal is a good one and that allows for
comparing proposals from event projects that had different goals. We
need a yardstick for history.

Our strategy for dealing with both problems is a plausibly objective
yardstick called a synthetic historical narrative. For both case
studies, we collect a large diverse corpus of narratives spanning
timelines, encyclopedia entries, journal articles, news reports,
websites, and government documents. Using natural language processing
(fully described in SI Appendix, Algorithm C1), we identify details that
appear across multiple accounts. The more accounts that mention a
detail, the more central it is to understanding the true historical
episode. The theoretical motivation is that authors face word limits
which force them to pick and choose which details to include, and they
choose details which serve the specific context of the document they are
producing. With a sufficiently large and diverse corpus of documents, we
can vary the context while holding the overall episode constant and see
which details tend to be invariant to context. Intuitively, a high
quality event dataset should have high recall for context invariant
details both because of their broader relevance and also because they
are easier to find in source material.

Synthetic historical narratives for Cuban Missiles (51 events drawn from
2020 documents) and Crimea-Donbas (30 events drawn from 971 documents)
appear in Figure 2. Each row represents a detail which appeared in at
least five documents along with an approximate start date, a hand
written summary, the number of documents it was mentioned in, and
whether it could be identified in the text of our ICB corpus, in our
ICBe events, or any of the competing systems.

From them, we draw several stylized facts. First, there is substantial
variation in which details any one document will choose to include. Our
ground truth ICB narratives included 17/51 and 23/30 of the events from
the synthetic narrative, while including other details that are not in
the synthetic narrative. Second, mentions of a detail across accounts is
exponentially distributed with context invariant details appearing
dozens to hundreds of times more than context dependent details. Third,
crisis start and stop dates are arbitrary and the historical record
points to many precursor events as necessary detail for understanding
later events, e.g.~the U.S. was in a \textit{de facto} grey scale war
with Cuba before it invited Soviet military protection (61) and Ukraine
provided several security guarantees to Russia that were potentially
undone, e.g.~a long term lease on naval facilities in Crimea. Fourth, we
find variation between the two cases. Cuban Missiles has a cleaner
canonical end with the Soviets agreeing to withdraw missiles while
Crimea-Donbas meekly ends with a second cease fire agreement (Minsk II)
but continued fighting. The canonical narrative of Cuban Missile also
includes high level previously classified details, while the more recent
Crimea-Donbas case reflects primarily public reporting.

We find substantive variation in recall across systems. Recall for each
increases in the number of document mentions which is an important sign
of validity for both them and our benchmark. The one outlier is Phoenix
which is so noisy that it's flat to decreasing in mentions. The two
episode level datasets have very low coverage of contextual details. The
two other dictionary systems ICEWs and Terrier have high coverage, with
ICEWs outperforming Terrier. ICBe strictly dominates all of the systems
but ICEWs in recall though we note that the small sample sizes mean
these systems should be considered statistically indistinguishable.
Importantly our corpus of ICB narratives has very high recall of
frequently mentioned details giving us confidence in how those summaries
were constructed, and ICBe lags only slightly behind showing that it
left very little additional information on the table.

\clearpage
\onecolumn
\begin{figure}[H]
\caption{Measuring Recall with Synthetic Historical Narratives \label{fig:case_study_cuban_recall}}
\centering{\includegraphics[height=18cm]{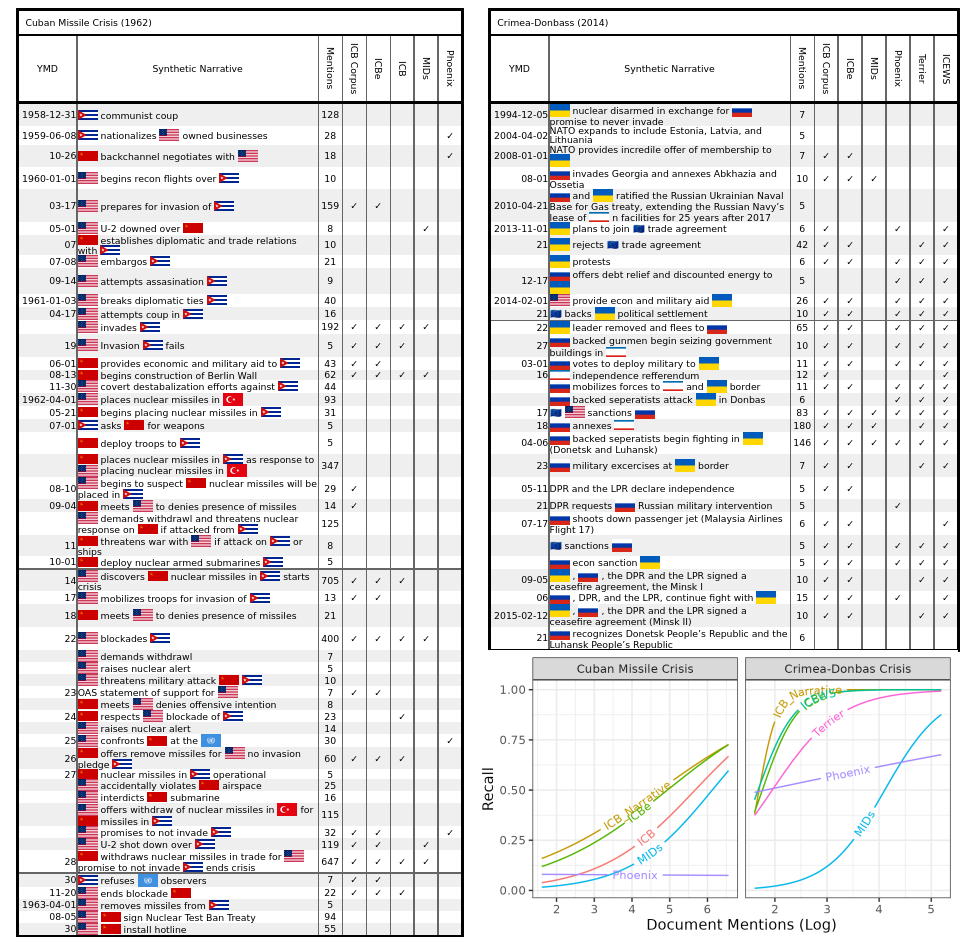}}
\textit{Notes: Synthetic narratives combine several thousand accounts of each crisis into a single timeline of events, taking only those mentioned in at least 5 or more documents. Checkmarks represent whether that event could be hand matched to any detail in the ICB corpus, ICBe dataset, or any of the other event datasets.}
\end{figure}
\clearpage
\twocolumn

\hypertarget{precision}{%
\section*{Precision}\label{precision}}
\addcontentsline{toc}{section}{Precision}

The other side of event measurement is precision, the degree to which a
sequence of events correctly and usefully describes the information in
history, \(Pr(H|E)\). It does little good to recall a historical event
but too vaguely (e.g.~MIDs describes the Cuban Missile crisis as a
blockade, a show of force, and a stalemate) or with too much error
(e.g.~ICEWS records 263 ``Detonate Nuclear Weapons'' events between
1995-2019) to be useful for downstream applications. ICBe's ontology and
coding system is designed to strike a balance so that the most important
information is recovered accurately but also abstracted to a level that
is still useful and interpretable. You should be able to lay out events
of a crisis on a timeline, as in Figure 3, and read off the macro
structure of an episode from each individual move. We call this
visualization a crisis map, a directed graph intersected with a
timeline, and provide crisis maps for every event dataset for each case
study (SI Appendix, Fig. D3 and D4) and all crises on the companion
website.

\clearpage
\onecolumn
\begin{sidewaysfigure}[ht]
\caption{Crisis Maps  \label{fig:p_precision_combined}}
\centering{\includegraphics[width=22cm]{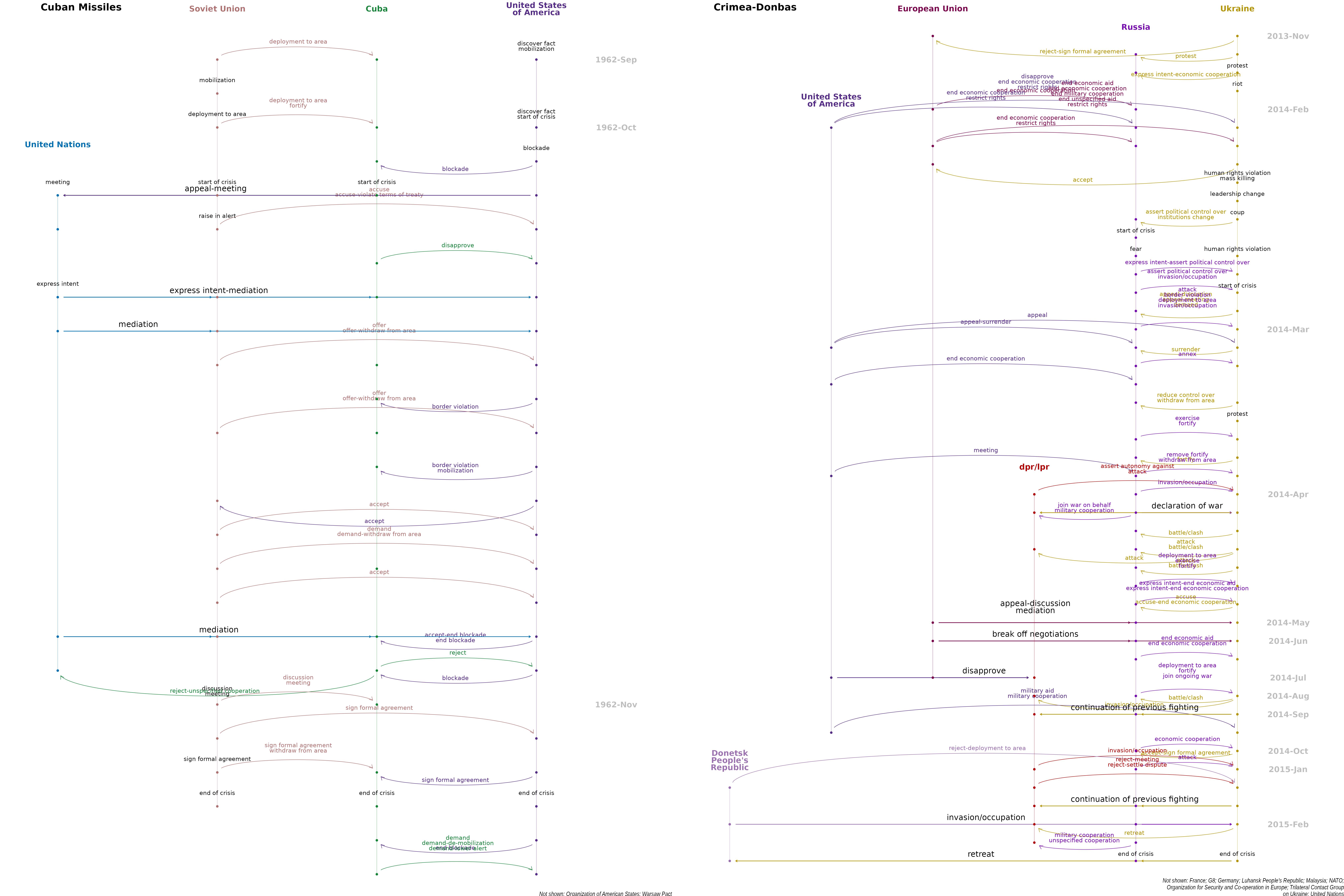}}
\end{sidewaysfigure}
\clearpage
\twocolumn

We further want to verify individual event codings, which we can do in
the case of ICBe because each event is mapped to a specific span of
text. We develop the iconography system for presenting event codings as
coherent statements that can be compared side by side to the original
source narrative as for Cuban Missiles (Figure 1), Crimea-Donbas (SI
Appendix Table D1), and for every case on the companion website. We
further provide a stratified sample of event codings alongside their
source text (SI Appendix Table D2).

We find both the visualizations of macro structure and head-to-head
comparisons of ICBe codings to the raw text to strongly support the
quality of ICBe, but as with recall we seek a more objective detached
universal benchmark. Our proposed measure is a reconstruction task to
see whether our intended ontology can be recovered through only
unsupervised clustering of sentences they were applied to. Figure 4
shows the location of every sentence from the ICBe corpus in semantic
space as embeded using the same large language model as before, and the
median location of each ICBe event tag applied to those
sentences.\footnote{We preprocess sentences to replace named entities
  with a generic Entity token.} Labels reflect the individual leaves of
the ontology and colors reflect the higher level coerce branch nodes of
the ontology. If ICBe has high precision, substantively similar tags
ought to have been applied to substantively similar source text, which
is what we see both in two dimensions in the main plot and via
hierarchical clustering on all dimensions in the dendrogram along the
righthand side.\footnote{Hierchcial clustering on cosine similarity and
  with Ward's method.}

\clearpage
\onecolumn
\begin{figure}[H]
\caption{ICBe event codings in comparison to Semantic Embeddings from source sentences\label{fig:semantic_embeddings}}
\centering{\includegraphics[width=17cm]{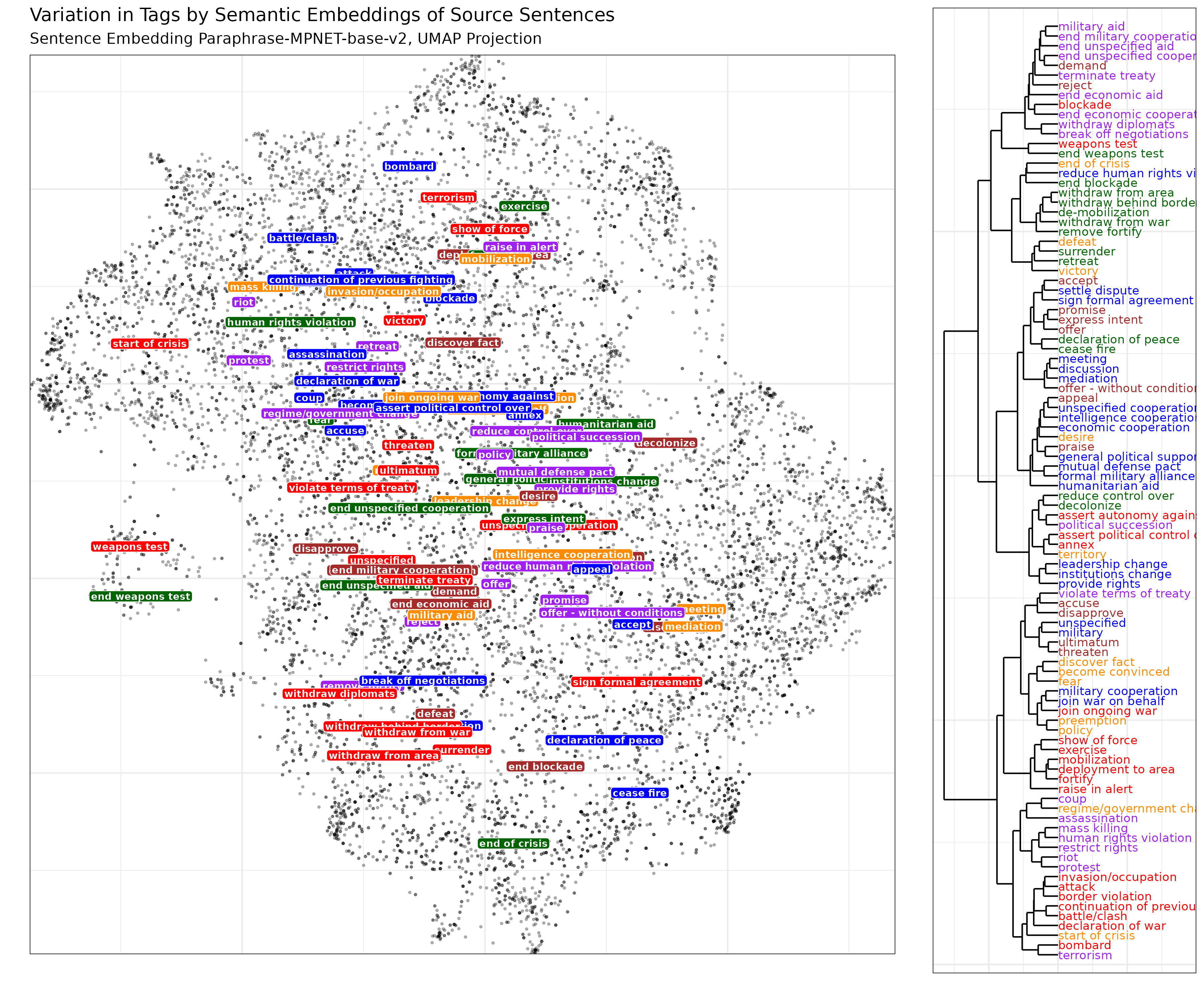}}
\textit{Notes: Dots represent individual ICB narrative sentences, as embeded by the Paraphrase-MPNET-base-v2 large language model and flattened into two dimensions with UMAP. Text labels reflect individual leaves of the ICBe ontology, and colors represent intermediate branches of the ontology. Label placement is the median of all of the sentences that tag was applied to by the coders. The dendrogram shows hiearchical clustering of the tags. If ICBe precision is high, the sentences tags were applied to ought to say similar things, and the intended shape of the ontology ought to be visually recognizable. }
\end{figure}
\clearpage
\twocolumn

Finally, how does ICBe's precision compare to the existing state of the
art? The crisis-maps reveal the episode level datasets like MIDs or the
original ICB are too sparse and vague to reconstruct the structure of
the crisis (SI Appendix Figure D3 and D4). On the other end of the
spectrum, the high recall dictionary based event datasets like Terrier
and ICEWs produce so many noisy events (several hundreds thousands) that
even with heavy filtering their crisis maps are completely
unintelligible. Further, because of copyright issues, none of these
datasets directly provide the original text spans making event level
precision difficult to verify.

However, given their high recall on our task and the global and
real-time coverage of dictionary based event systems, we want to take
seriously the possibility that some functional transformation could
recover the precision of ICBe. For example, (62) attempts to correct for
the mechanically increasing amount of news coverage each year by
detrending violent event counts from Phoenix using a human coded
baseline. Others have focused on verifying precision for ICEWs on
specific subsets of details against known ground truths,
e.g.~geolocation (63), protest events (80\%) (64), anti-government
protest networks (46.1\%) (65).

We take the same approach here in Figure 5, selecting four specific
CAMEO event codings and checking how often they reflect a true real
world event. We choose four event types around key moments in the
crisis. The start of the crisis revolves around Ukraine backing out of
trade deal with the EU in favor of Russia, but ``sign formal agreement''
events act more like a topic detector with dozens of events generated by
discussions of a possible agreement but not the actual agreement which
never materialized. The switch is caught by the ``reject plan, agreement
to settle dispute'', but also continues for Victor Yanukovych for even
after he was removed from power because of articles retroactively
discussing the cause of his of his removal. Events for ``use
conventional military force'' capture a threshold around the start of
hostilities and who the participants were but not any particular battles
or campaigns. Likewise, ``impose embargo, boycott, or sanctions''
captures the start of waves of sanctions and from who but are
effectively constantly as the news coverage does not distinguish between
subtle changes or additions. In sum, dictionary based methods on news
corpora tend to have high recall because they parse everything in the
news, but for the same reason their specificity for most event types is
too low to back out individual chess like sequencing that ICBe aims to
record.

\clearpage
\onecolumn
\begin{figure}[H]
\caption{ ICEWs Events by Day by Type during the Crimea-Donbas Crisis \label{fig:p_precision_icews}}
\centering{\includegraphics[width=17cm]{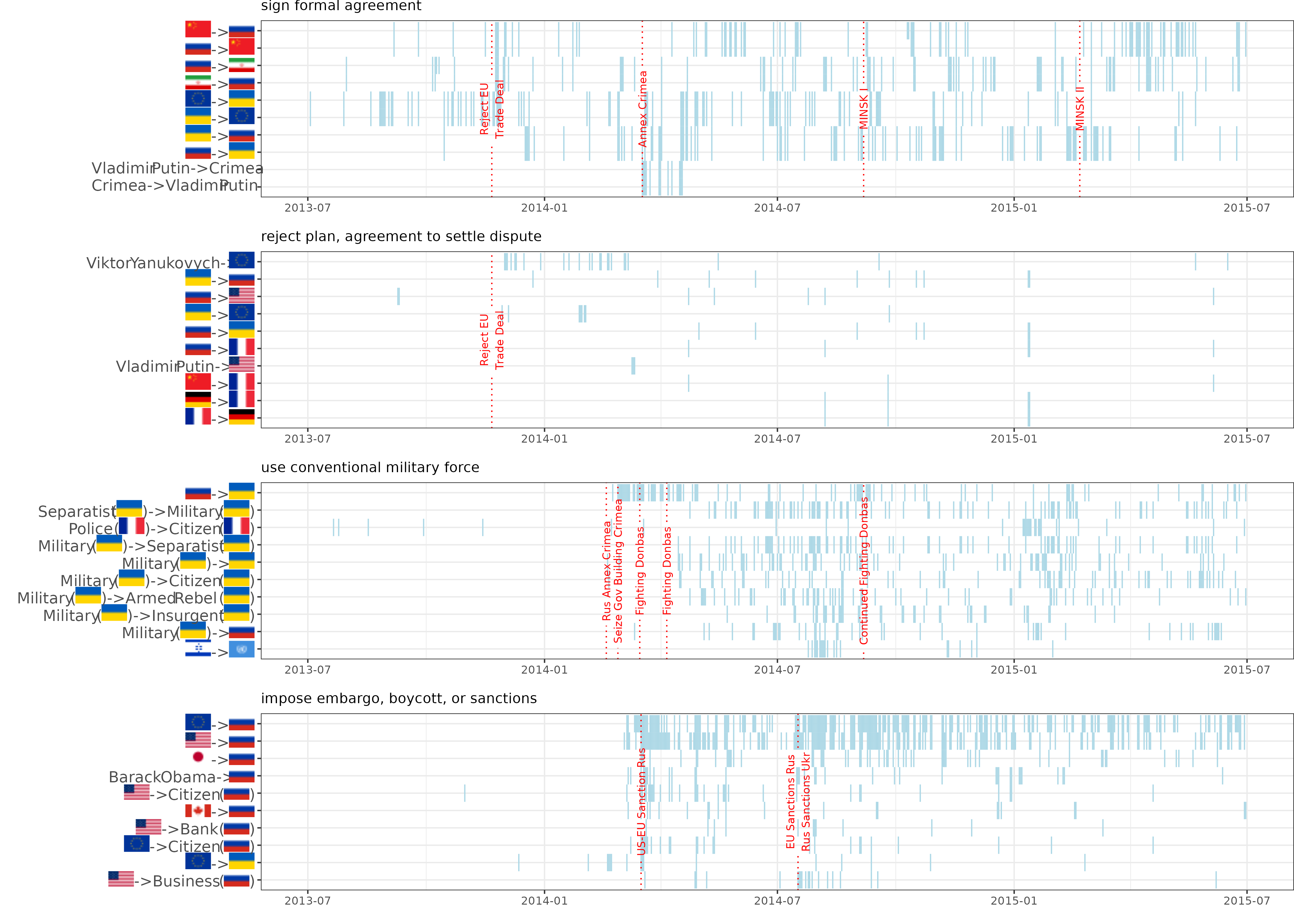}}
\textit{Notes:Unit of analysis is the Dyad-Day. Edges <-> indicates undirected dyad and -> indicates directed dyad. Top 10 most active dyads per category shown. Red text shows events from the synthetic narrative relative to that event category. Blue bars indicate an event recorded by ICEWs for that dyad on that day. }
\end{figure}
\clearpage
\twocolumn

\hypertarget{format}{%
\section*{Conclusion}\label{format}}
\addcontentsline{toc}{section}{Conclusion}

We investigated event abstraction from narratives describing key
historical episodes in international relations. We synthesized a prior
belief about the latent unobserved phenomena that drive these events in
international relations and proposed a mapping to observable concepts
that enter into the observed historical record. We designed an ontology
with high coverage over those concepts and developed a training
procedure and technical stack for human coding of historical texts.
Multiple validity checks find the resulting codings have high internal
validity (e.g.~intercoder agreement) and external validity
(i.e.~matching source material in both micro-details at the sentence
level and macro-details spanning full historical episodes). Further,
these codings perform much better in terms of recall, precision,
coverage, and overall coherence in capturing these historical episodes
than existing event systems used in international relations.

We release several open-source products along with supporting code and
documentation to further advance the study of IR, event extraction, and
natural language processing. The first is the International Crisis
Behavior Events (ICBe) dataset, an event-level aggregation of what took
place during the crises identified by the ICB project. These data are
appropriate for statistical analysis of hard questions about the
sequencing of events (e.g.~escalation and de-escalation of conflicts).
Second, we provide a coder-level disaggregation with multiple codings of
each sentence by experts and undergrads that allows for the introduction
of uncertainty and human interpretation of events. Further, we release a
direct mapping from the codings to the source text at the sentence level
as a new resource for natural language processing. Finally, we provide a
companion website that incorporates detailed visualizations of all of
the data introduced here (www.crisisevents.org).

\hypertarget{format}{%
\section*{Acknowledgements}\label{format}}
\addcontentsline{toc}{section}{Acknowledgements}

We thank the ICB Project and its directors and contributors for their
foundational work and their help with this effort. We make special
acknowledgment of Michael Brecher for helping found the ICB project in
1975, creating a resource that continues to spark new insights to this
day. We thank the many undergraduate coders for their patience and
dedication. Thanks to the Center for Peace and Security Studies and its
membership for comments. Special thanks to Rebecca Cordell, Philip
Schrodt, Zachary Steinert-Threlkeld, and Zhanna Terechshenko for
generous feedback. Thank you to the cPASS research assistants that
contributed to this project: Helen Chung, Daman Heer, Syeda ShahBano
Ijaz, Anthony Limon, Erin Ling, Ari Michelson, Prithviraj Pahwa, Gianna
Pedro, Tobias Stodiek, Yiyi `Effie' Sun, Erin Werner, Lisa Yen, and
Ruixuan Zhang. This project was supported by a grant from the Office of
Naval Research {[}N00014-19-1-2491{]} and benefited from the Charles
Koch Foundation's support for the Center for Peace and Security Studies.

\hypertarget{format}{%
\section*{Author Contributions}\label{format}}
\addcontentsline{toc}{section}{Author Contributions}

Conceptualization: R.W.D., E.G., J.L.; Methodology: R.W.D., T.L.S.;
Software: R.W.D.; Validation: R.W.D., T.L.S.; Formal Analysis: R.W.D.,
T.L.S.; Investigation: S.C., R.W.D., J.A.G., C.K., N.L., E.M., J.M.C.N.,
D.P., D.Q., J.W.; Data Curation: R.W.D., D.Q., T.L.S., J.W.; Writing -
Original Draft: R.W.D., T.L.S.; Writing - Review \& Editing: R.W.D.,
J.A.G., E.G., T.L.S.; Visualization: R.W.D., T.L.S.; Supervision: E.G.;
Project Administration: S.C., R.W.D., J.A.G., D.Q., T.L.S., J.W.;
Funding Acquisition: E.G., J.L.

~

\hypertarget{refs}{}
\begin{CSLReferences}{0}{0}
\leavevmode\vadjust pre{\hypertarget{ref-brecherInternationalStudiesTwentieth1999}{}}%
\CSLLeftMargin{1. }%
\CSLRightInline{Brecher M (1999) International studies in the twentieth
century and beyond: {Flawed} dichotomies, synthesis, cumulation: {ISA}
presidential address. \emph{International Studies Quarterly}
43(2):213--264.}

\leavevmode\vadjust pre{\hypertarget{ref-reiterShouldWeLeave2015}{}}%
\CSLLeftMargin{2. }%
\CSLRightInline{Reiter D (2015)
\href{https://doi.org/10.1146/annurev-polisci-053013-041156}{Should {We
Leave Behind} the {Subfield} of {International Relations}?} \emph{Annu
Rev Polit Sci} 18(1):481--499.}

\leavevmode\vadjust pre{\hypertarget{ref-brecherCrisesWorldPolitics1982}{}}%
\CSLLeftMargin{3. }%
\CSLRightInline{Brecher M, Wilkenfeld J (1982)
\href{https://doi.org/10.2307/2010324}{Crises in {World Politics}}.
\emph{World Politics} 34(3):380--417.}

\leavevmode\vadjust pre{\hypertarget{ref-beardsleyInternationalCrisisBehavior2020}{}}%
\CSLLeftMargin{4. }%
\CSLRightInline{Beardsley K, James P, Wilkenfeld J, Brecher M (2020) The
{International Crisis Behavior Project}.
doi:\href{https://doi.org/10.1093/acrefore/9780190228637.013.1638}{10.1093/acrefore/9780190228637.013.1638}.}

\leavevmode\vadjust pre{\hypertarget{ref-palmerMID5Dataset20112021}{}}%
\CSLLeftMargin{5. }%
\CSLRightInline{Palmer G, et al. (2021)
\href{https://doi.org/10.1177/0738894221995743}{The {MID5 Dataset},
2011--2014: {Procedures}, coding rules, and description}. \emph{Conflict
Management and Peace Science}:0738894221995743.}

\leavevmode\vadjust pre{\hypertarget{ref-giblerInternationalConflicts181620102018}{}}%
\CSLLeftMargin{6. }%
\CSLRightInline{Gibler DM (2018) \emph{International {Conflicts},
1816-2010: {Militarized Interstate Dispute Narratives}} ({Rowman \&
Littlefield}) Available at:
\url{https://books.google.com?id=_4VTDwAAQBAJ}.}

\leavevmode\vadjust pre{\hypertarget{ref-maozDyadicMilitarizedInterstate2019}{}}%
\CSLLeftMargin{7. }%
\CSLRightInline{Maoz Z, Johnson PL, Kaplan J, Ogunkoya F, Shreve AP
(2019) \href{https://doi.org/10.1177/0022002718784158}{The {Dyadic
Militarized Interstate Disputes} ({MIDs}) {Dataset Version} 3.0:
{Logic}, {Characteristics}, and {Comparisons} to {Alternative
Datasets}}. \emph{Journal of Conflict Resolution} 63(3):811--835.}

\leavevmode\vadjust pre{\hypertarget{ref-sarkeesResortWar181620072010}{}}%
\CSLLeftMargin{8. }%
\CSLRightInline{Sarkees MR, Wayman F (2010) \emph{Resort to war:
1816-2007} ({CQ Press}).}

\leavevmode\vadjust pre{\hypertarget{ref-reiterRevisedLookInterstate2016}{}}%
\CSLLeftMargin{9. }%
\CSLRightInline{Reiter D, Stam AC, Horowitz MC (2016)
\href{https://doi.org/10.1177/0022002714553107}{A {Revised Look} at
{Interstate Wars}, 1816--2007}. \emph{Journal of Conflict Resolution}
60(5):956--976.}

\leavevmode\vadjust pre{\hypertarget{ref-ralphsundbergUCDPGEDCodebook2016}{}}%
\CSLLeftMargin{10. }%
\CSLRightInline{Ralph Sundberg, Mihai Croicu (2016) \emph{{UCDP GED
Codebook} version 5.0} ({Department of Peace and Conflict Research,
Uppsala University}).}

\leavevmode\vadjust pre{\hypertarget{ref-petterssonOrganizedViolence19892018}{}}%
\CSLLeftMargin{11. }%
\CSLRightInline{Pettersson T, Eck K (2018)
\href{https://doi.org/10.1177/0022343318784101}{Organized violence,
1989--2017}. \emph{Journal of Peace Research} 55(4):535--547.}

\leavevmode\vadjust pre{\hypertarget{ref-raleighIntroducingACLEDArmed2010}{}}%
\CSLLeftMargin{12. }%
\CSLRightInline{Raleigh C, Linke A, Hegre H, Karlsen J (2010)
Introducing {ACLED}: An armed conflict location and event dataset:
Special data feature. \emph{Journal of peace research} 47(5):651--660.}

\leavevmode\vadjust pre{\hypertarget{ref-felbermayrGlobalSanctionsData2020}{}}%
\CSLLeftMargin{13. }%
\CSLRightInline{Felbermayr G, Kirilakha A, Syropoulos C, Yalcin E, Yotov
YV (2020) \href{https://doi.org/10.1016/j.euroecorev.2020.103561}{The
global sanctions data base}. \emph{European Economic Review}
129:103561.}

\leavevmode\vadjust pre{\hypertarget{ref-barariDemocracyTradePolicy}{}}%
\CSLLeftMargin{14. }%
\CSLRightInline{Barari S, Kim IS Democracy and {Trade Policy} at the
{Product Level}: {Evidence} from a {New Tariff-line Dataset}. 16.}

\leavevmode\vadjust pre{\hypertarget{ref-kinneDefenseCooperationAgreement2020}{}}%
\CSLLeftMargin{15. }%
\CSLRightInline{Kinne BJ (2020)
\href{https://doi.org/10.1177/0022002719857796}{The {Defense Cooperation
Agreement Dataset} ({DCAD})}. \emph{Journal of Conflict Resolution}
64(4):729--755.}

\leavevmode\vadjust pre{\hypertarget{ref-owsiakInternationalBorderAgreements2018}{}}%
\CSLLeftMargin{16. }%
\CSLRightInline{Owsiak AP, Cuttner AK, Buck B (2018)
\href{https://doi.org/10.1177/0738894216646978}{The {International
Border Agreements Dataset}}. \emph{Conflict Management and Peace
Science} 35(5):559--576.}

\leavevmode\vadjust pre{\hypertarget{ref-vabulasCooperationAutonomyBuilding2021}{}}%
\CSLLeftMargin{17. }%
\CSLRightInline{Vabulas F, Snidal D (2021)
\href{https://doi.org/10.1177/0022343320943920}{Cooperation under
autonomy: {Building} and analyzing the {Informal Intergovernmental
Organizations} 2.0 dataset}. \emph{Journal of Peace Research}
58(4):859--869.}

\leavevmode\vadjust pre{\hypertarget{ref-frederickIssueCorrelatesWar2017}{}}%
\CSLLeftMargin{18. }%
\CSLRightInline{Frederick BA, Hensel PR, Macaulay C (2017)
\href{https://doi.org/10.1177/0022343316676311}{The {Issue Correlates}
of {War Territorial Claims Data}, 1816--20011}. \emph{Journal of Peace
Research} 54(1):99--108.}

\leavevmode\vadjust pre{\hypertarget{ref-moyerWhatAreDrivers2020}{}}%
\CSLLeftMargin{19. }%
\CSLRightInline{Moyer JD, Turner SD, Meisel CJ (2020)
\href{https://doi.org/10.1177/0022343320929740}{What are the drivers of
diplomacy? {Introducing} and testing new annual dyadic data measuring
diplomatic exchange}. \emph{Journal of Peace
Research}:0022343320929740.}

\leavevmode\vadjust pre{\hypertarget{ref-sechserMilitarizedCompellentThreats2011}{}}%
\CSLLeftMargin{20. }%
\CSLRightInline{Sechser TS (2011)
\href{https://doi.org/10.1177/0738894211413066}{Militarized {Compellent
Threats}, 1918--2001}. \emph{Conflict Management and Peace Science}
28(4):377--401.}

\leavevmode\vadjust pre{\hypertarget{ref-liComprehensiveSurveySchemabased2021}{}}%
\CSLLeftMargin{21. }%
\CSLRightInline{Li Q, et al. (2021) A {Comprehensive Survey} on
{Schema-based Event Extraction} with {Deep Learning}. Available at:
\url{http://arxiv.org/abs/2107.02126} {[}Accessed September 10,
2021{]}.}

\leavevmode\vadjust pre{\hypertarget{ref-haltermanExtractingPoliticalEvents2020}{}}%
\CSLLeftMargin{22. }%
\CSLRightInline{Halterman A (2020) Extracting {Political Events} from
{Text Using Syntax} and {Semantics}.}

\leavevmode\vadjust pre{\hypertarget{ref-brandtPhoenixRealTimeEvent2018}{}}%
\CSLLeftMargin{23. }%
\CSLRightInline{Brandt PT, DOrazio V, Holmes J, Khan L, Ng V (2018)
Phoenix {Real-Time Event Data}. Available at:
\url{http://eventdata.utdallas.edu}.}

\leavevmode\vadjust pre{\hypertarget{ref-boscheeICEWSCodedEvent2015}{}}%
\CSLLeftMargin{24. }%
\CSLRightInline{Boschee E, et al. (2015) {ICEWS} coded event data.
\emph{Harvard Dataverse} 12.}

\leavevmode\vadjust pre{\hypertarget{ref-hegreIntroducingUCDPCandidate2020}{}}%
\CSLLeftMargin{25. }%
\CSLRightInline{Hegre H, Croicu M, Eck K, Högbladh S (2020) Introducing
the {UCDP Candidate Events Dataset}. \emph{Research \& Politics}
7(3):2053168020935257.}

\leavevmode\vadjust pre{\hypertarget{ref-grantOUEventData2017}{}}%
\CSLLeftMargin{26. }%
\CSLRightInline{Grant C, Halterman A, Irvine J, Liang Y, Jabr K (2017)
{OU Event Data Project}. Available at: \url{https://osf.io/4m2u7/}
{[}Accessed September 1, 2021{]}.}

\leavevmode\vadjust pre{\hypertarget{ref-zhangCASMDeepLearningApproach2019}{}}%
\CSLLeftMargin{27. }%
\CSLRightInline{Zhang H, Pan J (2019)
\href{https://doi.org/10.1177/0081175019860244}{{CASM}: {A Deep-Learning
Approach} for {Identifying Collective Action Events} with {Text} and
{Image Data} from {Social Media}}. \emph{Sociological Methodology}
49(1):1--57.}

\leavevmode\vadjust pre{\hypertarget{ref-steinert-threlkeldFutureEventData2019}{}}%
\CSLLeftMargin{28. }%
\CSLRightInline{Steinert-Threlkeld ZC (2019)
\href{https://doi.org/10.1177/0081175019860238}{The {Future} of {Event
Data Is Images}}. \emph{Sociological Methodology} 49(1):68--75.}

\leavevmode\vadjust pre{\hypertarget{ref-brecherCrisisEscalationWar2000}{}}%
\CSLLeftMargin{29. }%
\CSLRightInline{Brecher M, James P, Wilkenfeld J (2000) Crisis
escalation to war: {Findings} from the {International Crisis Behavior
Project}. \emph{What Do We Know About War}.}

\leavevmode\vadjust pre{\hypertarget{ref-wilkenfeldInterstateCrisesViolence2000}{}}%
\CSLLeftMargin{30. }%
\CSLRightInline{Wilkenfeld J, Brecher M (2000) Interstate crises and
violence: Twentieth-century findings. \emph{Handbook of war studies
II}:282--300.}

\leavevmode\vadjust pre{\hypertarget{ref-jamesWhatWeKnow2019}{}}%
\CSLLeftMargin{31. }%
\CSLRightInline{James P (2019)
\href{https://doi.org/10.1177/0738894218793135}{What do we know about
crisis, escalation and war? {A} visual assessment of the {International
Crisis Behavior Project}}. \emph{Conflict Management and Peace Science}
36(1):3--19.}

\leavevmode\vadjust pre{\hypertarget{ref-iakhnisCrisesWorldPolitics2019}{}}%
\CSLLeftMargin{32. }%
\CSLRightInline{Iakhnis E, James P (2019)
\href{https://doi.org/10.1177/0738894219855610}{Near crises in world
politics: {A} new dataset}. \emph{Conflict Management and Peace
Science}:0738894219855610.}

\leavevmode\vadjust pre{\hypertarget{ref-brecherInternationalCrisisBehavior2017}{}}%
\CSLLeftMargin{33. }%
\CSLRightInline{Brecher M, Wilkenfeld J, Beardsley KC, James P, Quinn D
(2017) \emph{International {Crisis Behavior Data Codebook}} Available
at: \url{http://sites.duke.edu/icbdata/data-collections/}.}

\leavevmode\vadjust pre{\hypertarget{ref-brecher_study_1997}{}}%
\CSLLeftMargin{34. }%
\CSLRightInline{Brecher M, Wilkenfeld J (1997) \emph{A {Study} of
{Crisis}} ({University of Michigan Press}).}

\leavevmode\vadjust pre{\hypertarget{ref-holsti1914Case1965}{}}%
\CSLLeftMargin{35. }%
\CSLRightInline{Holsti OR (1965)
\href{https://doi.org/10.2307/1953055}{The 1914 {Case}}. \emph{The
American Political Science Review} 59(2):365--378.}

\leavevmode\vadjust pre{\hypertarget{ref-paigeKoreanDecisionJune1968}{}}%
\CSLLeftMargin{36. }%
\CSLRightInline{Paige GD (1968) \emph{The {Korean Decision}, {June}
24-30, 1950} ({Free Press}).}

\leavevmode\vadjust pre{\hypertarget{ref-allisonEssenceDecisionExplaining1971}{}}%
\CSLLeftMargin{37. }%
\CSLRightInline{Allison GT, Zelikow P (1971) \emph{Essence of decision:
{Explaining} the {Cuban} missile crisis} ({Little, Brown Boston}).}

\leavevmode\vadjust pre{\hypertarget{ref-snyderConflictNationsBargaining1977}{}}%
\CSLLeftMargin{38. }%
\CSLRightInline{Snyder GH, Diesing P (1977) \emph{Conflict among
nations: {Bargaining} and decision making in international crises}
({Princeton University Press}).}

\leavevmode\vadjust pre{\hypertarget{ref-gavinHistorySecurityStudies2014}{}}%
\CSLLeftMargin{39. }%
\CSLRightInline{Gavin FJ (2014)
\href{https://doi.org/10.1080/01402390.2014.912916}{History, {Security
Studies}, and the {July Crisis}}. \emph{Journal of Strategic Studies}
37(2):319--331.}

\leavevmode\vadjust pre{\hypertarget{ref-georgeDeterrenceAmericanForeign1974}{}}%
\CSLLeftMargin{40. }%
\CSLRightInline{George AL, Smoke R (1974) \emph{Deterrence in {American}
foreign policy: {Theory} and practice} ({Columbia University Press}).}

\leavevmode\vadjust pre{\hypertarget{ref-gaddisExpandingDataBase1987}{}}%
\CSLLeftMargin{41. }%
\CSLRightInline{Gaddis JL (1987)
\href{https://doi.org/10.2307/2538915}{Expanding the {Data Base}:
{Historians}, {Political Scientists}, and the {Enrichment} of {Security
Studies}}. \emph{International Security} 12(1):3--21.}

\leavevmode\vadjust pre{\hypertarget{ref-brecherPatternsCrisisManagement1988}{}}%
\CSLLeftMargin{42. }%
\CSLRightInline{Brecher M, James P (1988) Patterns of crisis management.
\emph{Journal of Conflict Resolution} 32(3):426--456.}

\leavevmode\vadjust pre{\hypertarget{ref-kangUSBiasStudy2019}{}}%
\CSLLeftMargin{43. }%
\CSLRightInline{Kang DC, Lin AY-T (2019) {US} bias in the study of
{Asian} security: {Using Europe} to study {Asia}. \emph{Journal of
Global Security Studies} 4(3):393--401.}

\leavevmode\vadjust pre{\hypertarget{ref-hewittEngagingInternationalData2001}{}}%
\CSLLeftMargin{44. }%
\CSLRightInline{Hewitt JJ (2001)
\href{https://doi.org/10.1111/1528-3577.00066}{Engaging {International
Data} in the {Classroom}: {Using} the {ICB Interactive Data Library} to
{Teach Conflict} and {Crisis Analysis}}. \emph{Int Stud Perspect}
2(4):371--383.}

\leavevmode\vadjust pre{\hypertarget{ref-millerWordNetLexicalDatabase1995}{}}%
\CSLLeftMargin{45. }%
\CSLRightInline{Miller GA (1995)
\href{https://doi.org/10.1145/219717.219748}{{WordNet}: A lexical
database for {English}}. \emph{Commun ACM} 38(11):39--41.}

\leavevmode\vadjust pre{\hypertarget{ref-althausClineCenterHistorical2019}{}}%
\CSLLeftMargin{46. }%
\CSLRightInline{Althaus S, Bajjalieh J, Carter JF, Peyton B, Shalmon DA
(2019) Cline {Center Historical Phoenix Event Data Variable
Descriptions}. \emph{Cline Center Historical Phoenix Event Data}.}

\leavevmode\vadjust pre{\hypertarget{ref-sundbergIntroducingUCDPGeoreferenced2013}{}}%
\CSLLeftMargin{47. }%
\CSLRightInline{Sundberg R, Melander E (2013) Introducing the {UCDP}
georeferenced event dataset. \emph{Journal of Peace Research}
50(4):523--532.}

\leavevmode\vadjust pre{\hypertarget{ref-lengMilitarizedInterstateCrises1988}{}}%
\CSLLeftMargin{48. }%
\CSLRightInline{Leng RJ, Singer JD (1988)
\href{https://doi.org/10.2307/2600625}{Militarized {Interstate Crises}:
{The BCOW Typology} and {Its Applications}}. \emph{International Studies
Quarterly} 32(2):155--173.}

\leavevmode\vadjust pre{\hypertarget{ref-mcclellandWorldEventInteraction1978}{}}%
\CSLLeftMargin{49. }%
\CSLRightInline{McClelland C (1978) World event/interaction survey,
1966-1978. \emph{WEIS Codebook ICPSR} 5211.}

\leavevmode\vadjust pre{\hypertarget{ref-hermannComparativeResearchEvents1984}{}}%
\CSLLeftMargin{50. }%
\CSLRightInline{Hermann C (1984) Comparative {Research} on the {Events}
of {Nations} ({CREON}) {Project}: {Foreign Policy Events}, 1959-1968:
{Version} 1.
doi:\href{https://doi.org/10.3886/ICPSR05205.V1}{10.3886/ICPSR05205.V1}.}

\leavevmode\vadjust pre{\hypertarget{ref-bloomfieldCASCONIIIComputeraided1989}{}}%
\CSLLeftMargin{51. }%
\CSLRightInline{Bloomfield LP, Moulton A (1989) {CASCON III}:
{Computer-aided} system for analysis of local conflicts. \emph{MIT
Center for International Studies, Cambridge}.}

\leavevmode\vadjust pre{\hypertarget{ref-shermanSHERFACSCrossParadigmHierarchical2000}{}}%
\CSLLeftMargin{52. }%
\CSLRightInline{Sherman FL (2000) {SHERFACS}: {A Cross-Paradigm},
{Hierarchical}, and {Contextually-Sensitive International Conflict
Dataset}, 1937-1985: {Version} 1.
doi:\href{https://doi.org/10.3886/ICPSR02292.V1}{10.3886/ICPSR02292.V1}.}

\leavevmode\vadjust pre{\hypertarget{ref-balaliCOfEEComprehensiveOntology2021}{}}%
\CSLLeftMargin{53. }%
\CSLRightInline{Balali A, Asadpour M, Jafari SH (2021) {COfEE}: {A
Comprehensive Ontology} for {Event Extraction} from text.
doi:\href{https://doi.org/10.48550/arXiv.2107.10326}{10.48550/arXiv.2107.10326}.}

\leavevmode\vadjust pre{\hypertarget{ref-merrittMeasuringEventsInternational1994}{}}%
\CSLLeftMargin{54. }%
\CSLRightInline{Merritt RL (1994) Measuring events for international
political analysis. \emph{International Interactions} 20(1-2):3--33.}

\leavevmode\vadjust pre{\hypertarget{ref-schrodtTwentyYearsKansas2006}{}}%
\CSLLeftMargin{55. }%
\CSLRightInline{Schrodt PA, Hall B (2006) Twenty years of the {Kansas}
event data system project. \emph{The political methodologist}
14(1):2--8.}

\leavevmode\vadjust pre{\hypertarget{ref-brecherInternationalCrisisBehavior}{}}%
\CSLLeftMargin{56. }%
\CSLRightInline{Brecher M, Wilkenfeld J, Beardsley K, James P, Quinn D
International {Crisis Behavior Data Codebook}, {Version} 12. 69.}

\leavevmode\vadjust pre{\hypertarget{ref-braithwaiteMIDLOCIntroducingMilitarized2010}{}}%
\CSLLeftMargin{57. }%
\CSLRightInline{Braithwaite A (2010)
\href{https://doi.org/10.1177/0022343309350008}{{MIDLOC}: {Introducing}
the {Militarized Interstate Dispute Location} dataset}. \emph{Journal of
Peace Research} 47(1):91--98.}

\leavevmode\vadjust pre{\hypertarget{ref-braithwaiteCodebookMilitarizedInterstate2009}{}}%
\CSLLeftMargin{58. }%
\CSLRightInline{Braithwaite A (2009) Codebook for the {Militarized
Interstate Dispute Location} ({MIDLOC}) {Data}, v 1.0. \emph{University
College London}.}

\leavevmode\vadjust pre{\hypertarget{ref-brustIntegratingDomainKnowledge2020}{}}%
\CSLLeftMargin{59. }%
\CSLRightInline{Brust C-A, Denzler J (2020) Integrating domain
knowledge: Using hierarchies to improve deep classifiers. Available at:
\url{http://arxiv.org/abs/1811.07125} {[}Accessed September 7, 2021{]}.}

\leavevmode\vadjust pre{\hypertarget{ref-mandelbrotFractalGeometryNature1983}{}}%
\CSLLeftMargin{60. }%
\CSLRightInline{Mandelbrot BB (1983) \emph{The fractal geometry of
nature} ({Freeman}, {New York}).}

\leavevmode\vadjust pre{\hypertarget{ref-cormacGreyNewBlack2018}{}}%
\CSLLeftMargin{61. }%
\CSLRightInline{Cormac R, Aldrich RJ (2018)
\href{https://doi.org/10.1093/ia/iiy067}{Grey is the new black: Covert
action and implausible deniability}. \emph{International Affairs}
94(3):477--494.}

\leavevmode\vadjust pre{\hypertarget{ref-terechshenkoHotCollarLatent2020}{}}%
\CSLLeftMargin{62. }%
\CSLRightInline{Terechshenko Z (2020)
\href{https://doi.org/10.1177/0022343320962546}{Hot under the collar:
{A} latent measure of interstate hostility}. \emph{Journal of Peace
Research} 57(6):764--776.}

\leavevmode\vadjust pre{\hypertarget{ref-cookLostAggregationImproving2019}{}}%
\CSLLeftMargin{63. }%
\CSLRightInline{Cook SJ, Weidmann NB (2019)
\href{https://doi.org/10.1111/ajps.12398}{Lost in {Aggregation}:
{Improving Event Analysis} with {Report-Level Data}}. \emph{American
Journal of Political Science} 63(1):250--264.}

\leavevmode\vadjust pre{\hypertarget{ref-wuestExternalValidationProtest2020}{}}%
\CSLLeftMargin{64. }%
\CSLRightInline{Wüest B, Lorenzini J (2020) External validation of
protest event analysis. \emph{Contention in times of crisis: Recession
and political protest in thirty Euro-pean countries}:49--78.}

\leavevmode\vadjust pre{\hypertarget{ref-jagerLimitsStudyingNetworks}{}}%
\CSLLeftMargin{65. }%
\CSLRightInline{Jäger K The {Limits} of {Studying Networks} with {Event
Data}: {Evidence} from the {ICEWS Dataset}.}

\end{CSLReferences}

\bibliographystyle{unsrt}
\bibliography{paper.bib}

\end{document}